\documentclass[a4paper,10pt]{article}

\usepackage{contribution}
\newcommand{\weblink}[2][]{%
    \ifthenelse{\equal{#1}{}}%
    {\textnormal{\url{#2}}}%
    {\textnormal{\href{#2}{#1}}}%
}

\def\beq{\begin{equation}}
\def\eeq#1{\label{#1}\end{equation}}
\def\eeqn{\end{equation}}

\def\beqa{\begin{eqnarray}}
\def\eeqa#1{\label{#1}\end{eqnarray}}
\def\eeqan{\end{eqnarray}}

\let\bar=\overbar

\def\One{{\bf 1}}

\def\Dslash{\not{\hbox{\kern-4pt $D$}}}
\def\dslash{\not{\hbox{\kern-2pt $\del$}}}

\def\msb{{\bar{\ssstyle M \kern -1pt S}}}

\newcommand{\contribution}[7][]{%
  \clearpage
  \thispagestyle{plain}

  \ifthenelse{\equal{#1}{}}
  {\hypersetup{pdftitle={#2}}}
  {\hypersetup{pdftitle={#1}}}
  \hypersetup{pdfauthor={{#3} {#4}}}
  {\centering\normalfont\LARGE\bfseries\sffamily #2 \par\nobreak}
  \lhead{}
  \chead{%
    \textit{\footnotesize XXIInd International Workshop ``High-Energy Physics and Quantum Field Theory'', 
June 24 -- July 1, 2015, Samara, Russia}%
  }
  \rhead{}
  \bigskip
  \begin{center}
    {#3} {#4}\ifthenelse{\equal{#6}{}}{}{\footnote{\weblink[#6]{mailto:#6}}}
    \ifthenelse{\equal{#7}{}}{}{#7} \\
    \textit{#5}
  \end{center}
  \bigskip
}

\renewcommand{\abstract}[1]{%
  \begin{center}
    \begin{minipage}{0.85\textwidth}
      \begin{footnotesize}
        #1
      \end{footnotesize}
    \end{minipage}
  \end{center}
  \bigskip
}

\begin{document} 

% % % % % % % % % % % % % % % % % % % % % % % % % % % % % % % % % % % % % % % % %
% your proceedings
% template for qfthep2015 contribution
%
% please do not rename this file
%
% to create document run
%
%     pdflatex qfthep (if you have pdf figures)
%
%     or 
%     latex qfthep
%     dvips qfthep
%     ps2pdf qfthep.ps
%
%     Finally, please rename your contribution qfthep.pdf -> <your name>.pdf 
%     and send us <your name>.pdf file
%
%%%%%%%%%%%%%%%%%%%%%%%%%%%%%%%%%%%%%%%%%%%%%%%%%%%%%%%%%%%%%%%%%%%%%%%%%%%%%%%%%
{  % do not remove
%%%%%%%%%%%%%%%%%%%%%%%%%%%%%%%%%%%%%%%%%%%%%%%%%%%%%%%%%%%%%%%%%%%%%%%%%%%%%%%%%
% template for articles submitted to the full-tex econf proceedings

%%%%%%%%%%%%%%%%%%%%%%%%%%%%%%%%%%%%%%%%%%%%%%%%%%%%%%%%%%%%%%%%%%%%%%%%%%%%%%%%%
% please define your macros here

%
%%%%%%%%%%%%%%%%%%%%%%%%%%%%%%%%%%%%%%%%%%%%%%%%%%%%%%%%%%%%%%%%%%%%%%%%%%%%%%%%%

%%%%%%%%%%%%%%%%%%%%%%%%%%%%%%%%%%%%%%%%%%%%%%%%%%%%%%%%%%%%%%%%%%%%%%%%%%%%%%%%%
% define title, author, and address
% contribution[short title]{title}{author first name}{author last name}{author address}{author email}{collaboration}
% the short title will appear in the page headers and the TOC of the book of proceedings
% the last two arguments may be left empty
\contribution[Proton and kaon timelike form factors from BABAR]  % short title (optional)
{Proton and kaon timelike form factors from BABAR}  % title
{S.~I.}{Serednyakov \\}  % first and last name of author
{Novosibirsk State University, \\
Budker Institute of Nuclear Physics \\
630090 Novosibirsk, Russia}  % author address
{seredn@inp.nsk.su}  % author email optional dmitri\_melikhov@gmx.de
{(on behalf of BABAR collaboration)}  % collaboration (optional, co-authors appear here )
%
%%%%%%%%%%%%%%%%%%%%%%%%%%%%%%%%%%%%%%%%%%%%%%%%%%%%%%%%%%%%%%%%%%%%%%%%%%%%%%%%%

%%%%%%%%%%%%%%%%%%%%%%%%%%%%%%%%%%%%%%%%%%%%%%%%%%%%%%%%%%%%%%%%%%%%%%%%%%%%%%%%%
% abstract
\abstract{%
The latest BABAR results on the proton and kaon timelike form factors (FF)
are presented. The special emphasize is made on  comparison of the spacelike 
and timelike FFs and the rise of the proton FF near threshold. The behavior of
the cross section of  e+e- annihilation into hadrons  near the nucleon-antinucleon
threshold is discussed.  }
%
%%%%%%%%%%%%%%%%%%%%%%%%%%%%%%%%%%%%%%%%%%%%%%%%%%%%%%%%%%%%%%%%%%%%%%%%%%%%%%%%%

%%%%%%%%%%%%%%%%%%%%%%%%%%%%%%%%%%%%%%%%%%%%%%%%%%%%%%%%%%%%%%%%%%%%%%%%%%%%%%%%%
% main text
% for short contributions sections are optional

\section{Introduction} 
The cross sections of the $e^+e^-$ annihilation into hadrons are described in 
terms of the electromagnetic 
form factors (FF). In case of production of proton-antiproton 
pair 
\begin{equation}
e^+e^- \to p \bar{p}
\label{eqB1}
\end{equation}
the cross section depends on two such functions, electric ($G_E$) and
magnetic ($G_M$) FFs:
\begin{equation}
\sigma_{p\bar{p}}(s) = \frac{4\pi\alpha^{2}\beta C}{3s}
\left [|G_M(s)|^{2} + \frac{1}{2\tau}|G_E(s)|^{2}\right]
\label{eqB2}
\end{equation}
where $s$ is the $e^+e^-$ center-of-mass (c.m.) energy squared,
$\beta = \sqrt{1-4m_B^2/s}$,
$C$ is the Coulomb interaction factor
[$C = y/(1-e^{-y})$ with $y = {\pi\alpha (1+\beta^2)}/{\beta }$
for protons, and $C=1$ for neutrons], $\tau = s/4m_B^2$. 

From the measurement of the total cross section the linear combination
of squared form factors
\begin{equation}
F(s)^2=\frac{2\tau |G_M(s)|^2+|G_E(s)|^2}{2\tau +1}
\label{eqB22}
\end{equation}
can be determined. The function $F(s)$ is called the effective form factor.
It is  this function that is measured in most of experiments.

%\noindent
         In case of production of kaons pair
\begin{equation}
e^+e^- \to K^+ K^-
\label{eqB3}
\end{equation}
the expression for the cross section has the following form:
\begin{equation}
\sigma_{K\bar{K}}(s) = \frac{\pi\alpha^{2}\beta^3}{3s} |F_K(s)|^2
\label{eqB4}
\end{equation}

There are many models describing timelike (TL) FFs, but  definite
predictions exist only for asymptotic region $s\to\infty$:

\begin{equation}
G_{E,M}(s)=G_{E,M}(-s)\sim \alpha_s^2(s)/s^2,
\label{eqB5}
\end{equation}
\begin{equation}
F_K(s)=(8\pi f_K^2\alpha_s(s))/s
\label{eqB6}
\end{equation}
where $\alpha_s\sim 1/\ln (s/\Lambda^2)$ is the  strong coupling constant,
$f_K$=156 MeV is the  $K\to l\nu$ decay constant.

    In the BABAR experiment the initial state radiation
(ISR) method was developed and used to measure
$e^+e^- \to$ hadrons cross sections at energies lower than the 
collider (c.m.) energy. In this talk the latest results 
on the proton and charged kaon TL  FFs from BABAR  are presented.

\begin{figure}
\includegraphics[width=0.47\textwidth]{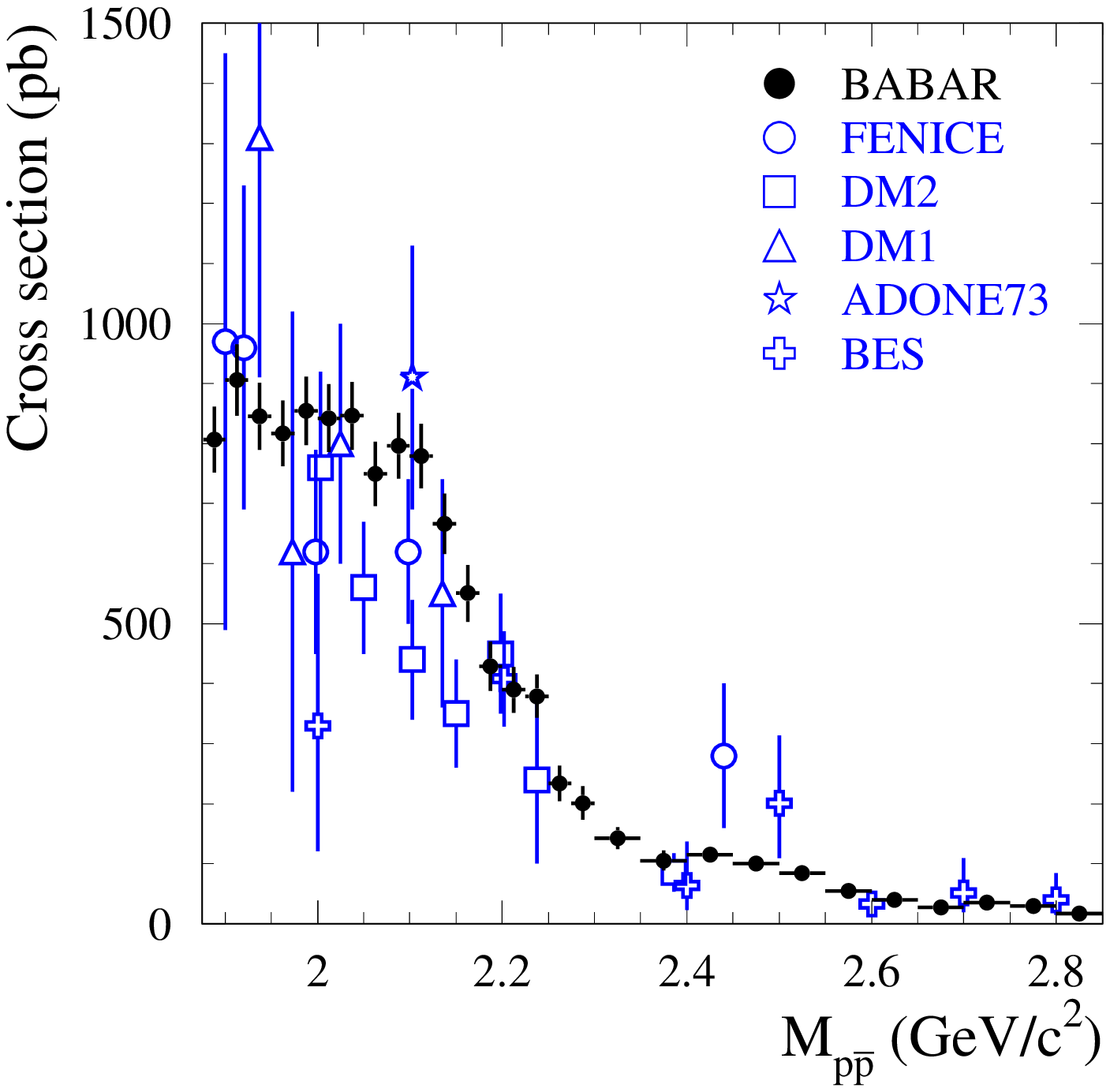}
\includegraphics[width=0.47\textwidth]{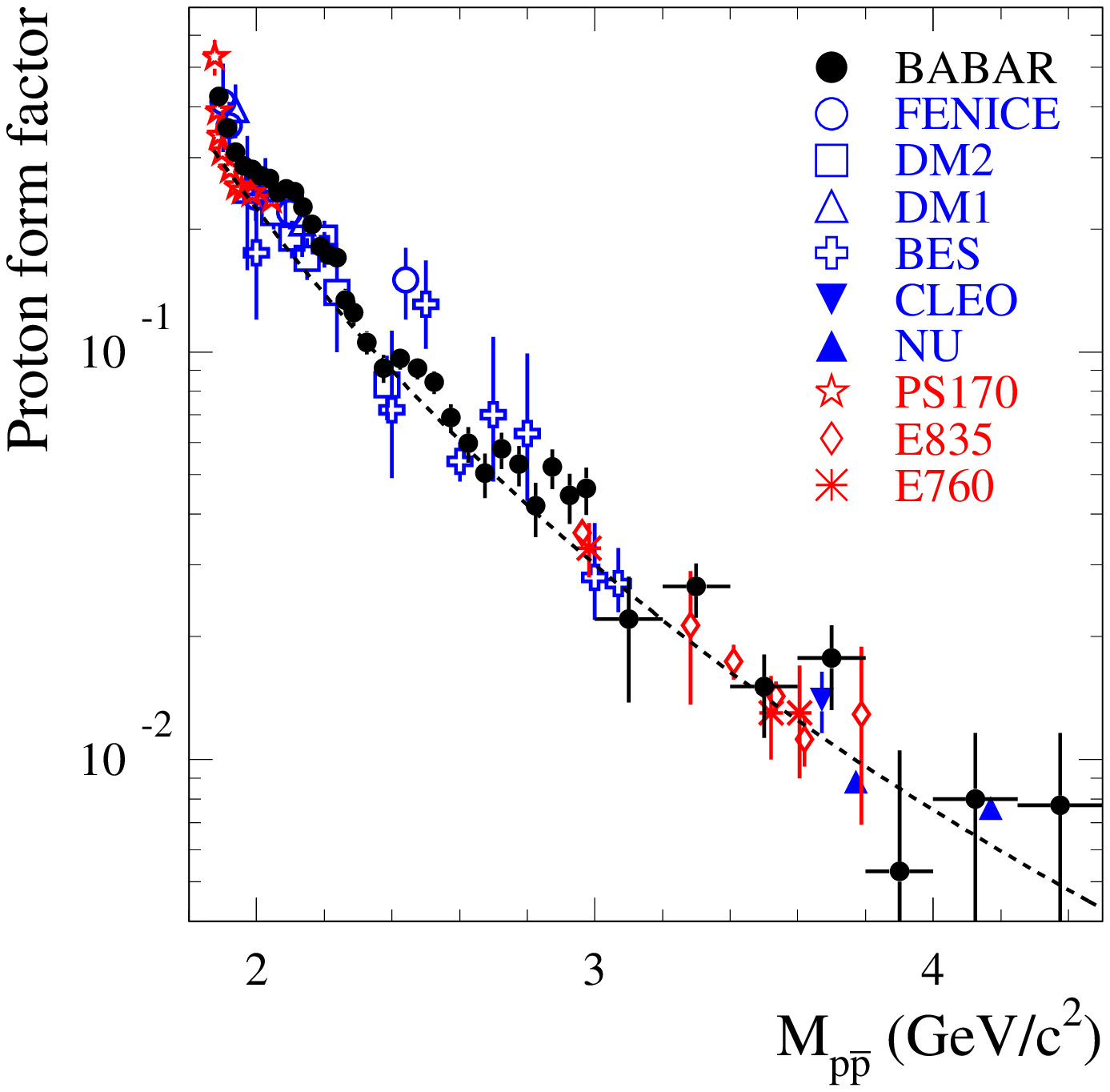}
\caption{ (Color online)  {\bf Left}:
The $e^+e^- \to p \bar{p}$ cross section near threshold
measured   by BABAR \cite{pp1} and in other experiments.
{\bf Right}:
The proton effective form factor [Eq.\ref{eqB22}]
measured by BABAR
\cite{pp1} and in other experiments. 
The curve is the QCD motivated fit [Eq.\ref{eqB5}].
\label{ppform1}}
\end{figure}

\section{ The proton form factor}
There are two BABAR experiments \cite{pp1,pp2} on the proton FF,
which use different ISR techniques. 
In the first method called large angle (LA) ISR, the ISR photon and  
proton-antiproton pair is required to be detected. LA ISR is effective at
lower masses $m_{p\bar{p}}<4~GeV/c^2$.
The cross section for  the  process (\ref{eqB1}) measured using LA ISR 
in the near threshold region \cite{pp1} is shown in Fig.\ref{ppform1} (left).
It slowly varies  from the threshold up to 2.1 GeV/c$^2$
and  then sharply goes down from 0.85 to  0.1 nb.
Such a behaviour of the cross section can be explained by the final state 
nucleon-antinucleon interaction \cite{Milst}. 
The proton FF value is close to 0.5 at the  threshold
(Fig.\ref{ppform1} (right)).  Then  it  decreases by two orders of magnitude
up to  4.5 GeV/c$^2$. 
Some deviations from the $s^{-2}$ fit seen at 2.15 GeV/c$^2$ and 
2.9 GeV/c$^2$ can be understood as  contributions of $p\bar{\Delta}$(1232) 
and $N$(1520)$\bar{N}$(1520) intermediate states respectively.

  In the  second  BABAR measurement of the proton FF,   
the ISR photon is required to be emitted at small angles  (SA ISR) 
and undetected.
The SA ISR is effective in the  $p\bar{p}$
mass range above 3 GeV/c$^2$. 

The proton FF  measured by BABAR using  the SA  ISR
\cite{pp2} is shown in  Fig.\ref{ppform2} (left). For  comparison,  the
spacelike (SL) proton FF data are shown. The TL and SL FFs should be equil
in the asymptotic limit.   Below 4 GeV/c$^2$
the TL  values are higher than the  SL  values by two times
(Fig.\ref{ppform2} (left)).
But beginning from 5 GeV/c$^2$ the tendency of approaching TL data to SL data is
seen. 

  As it seen in  the Fig.\ref{ppform1} (left) the $e^+e^-\to p\bar{p}$
cross section, measured by BABAR,
has a step-like shape with a step  height of  about 0.85 nb. 
The similar behaviour is observed for 
the $e^+e^-\to n\bar{n}$ cross section \cite{nnbar}.
The sum of these cross sections  shown in Fig.\ref{ppform2} (right,b)  has the
step height near  1.7 nb.   One can expect that such a  step 
%in the nucleon-antinucleon cross section  
must be compensated by a  similar negative 
step in the meson production cross section.
It was noticed in the work \cite{pi6NN} 	 
that such  1.7 nb negative step is observed in the
$e^+e^-\to 6\pi$ cross section. So,  the total
cross section (sum of $e^+e^-\to 6\pi$ and $e^+e^-\to p\bar{p},n\bar{n}$)
has no structure (Fig.\ref{ppform2} (right,c)). 
Today  there is no clear
understanding,  why only the $e^+e^-\to 6\pi$ process 
is sufficient to compensate
the nucleon-antinucleon step.
   
\begin{figure}
\includegraphics[width=0.47\textwidth]{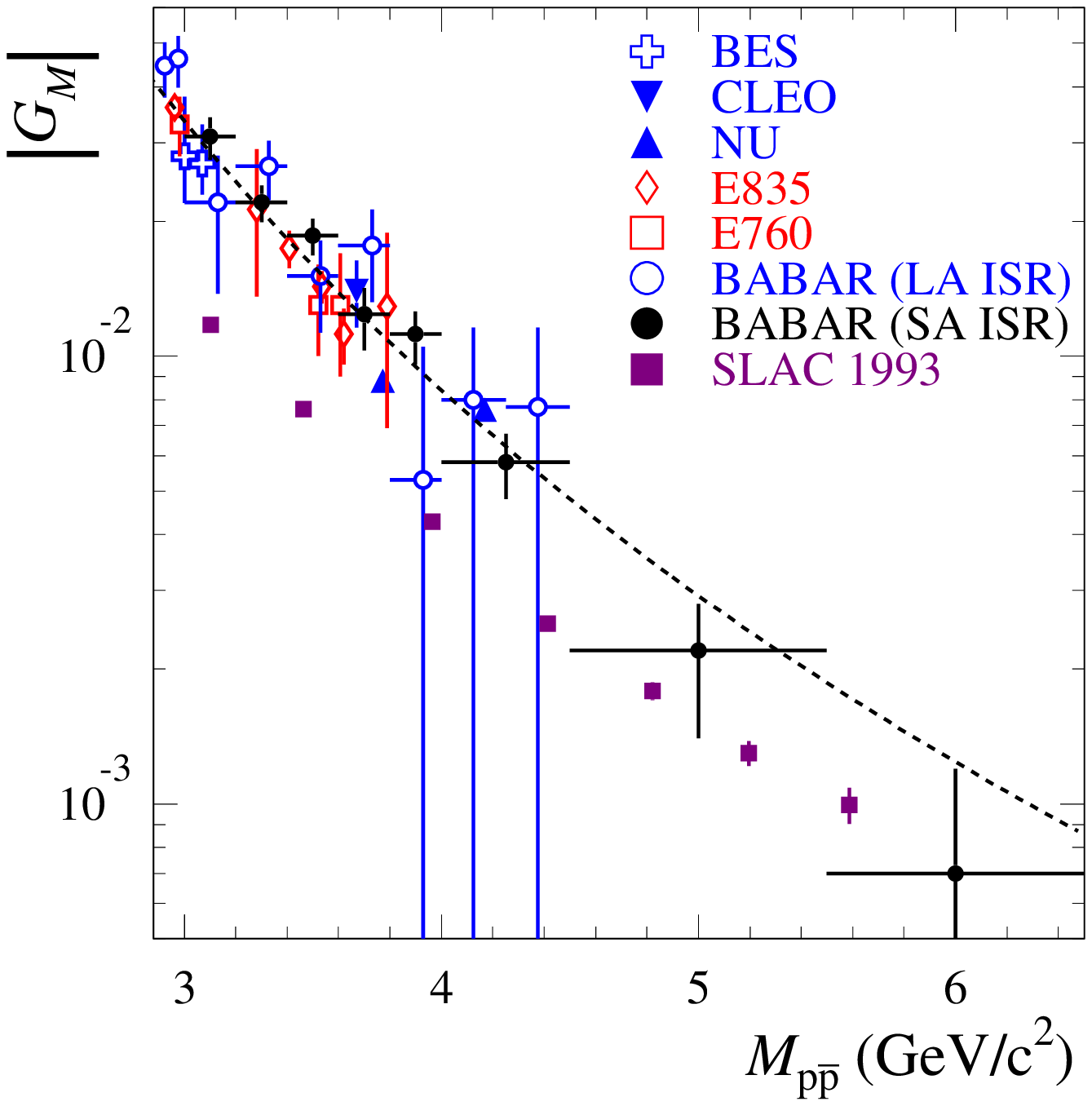}
\includegraphics[width=0.47\textwidth]{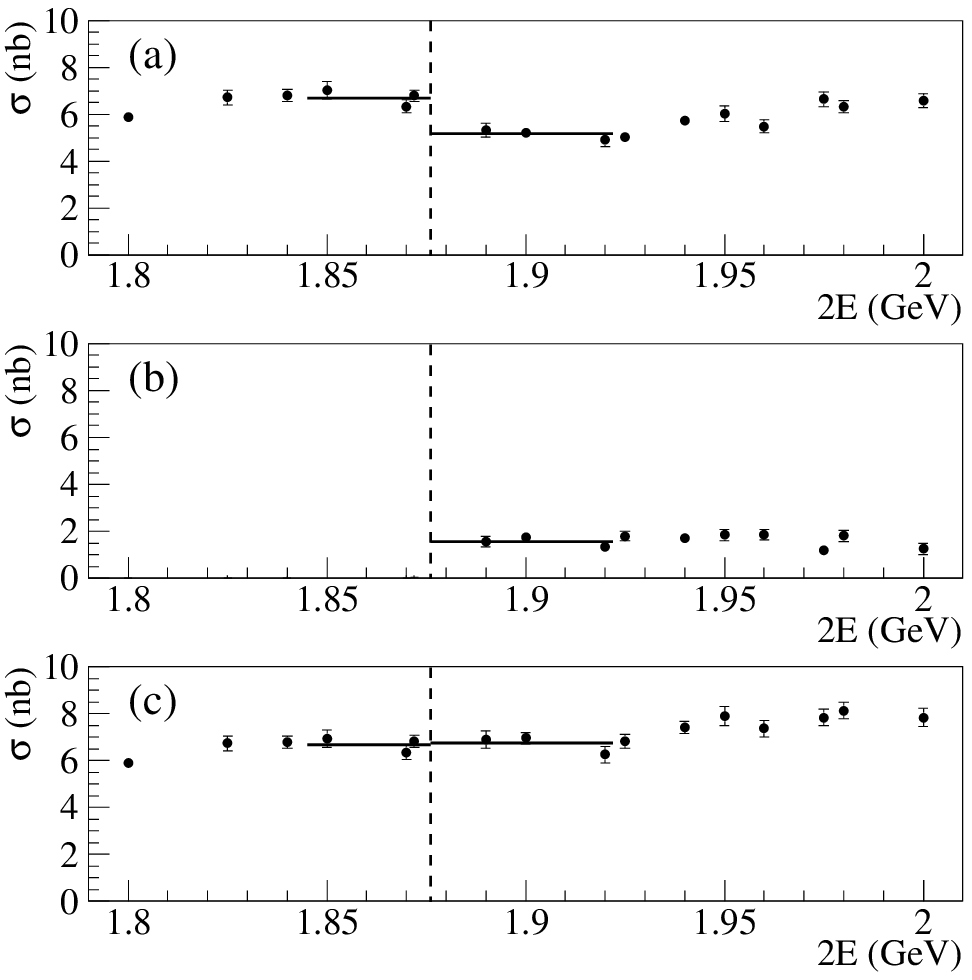}
\caption{{  (Color online)   \bf Left}: 
The proton magnetic form factor measured by BABAR
\cite{pp2} and in other experiments. The curve is the QCD fit.
The SLAC 1993 points  are the spacelike form
factor  data  from $ep$ scattering experiment.
{\bf Right}: The cross sections near nucleon-antinucleon threshold \cite{pi6NN}
for   $e^+e^-\to 6\pi$ (a), for   $e^+e^-\to p\bar{p},n\bar{n}$ (b), 
and for the sum of $p\bar{p},n\bar{n}$ and $6\pi$ processes  (c).
\label{ppform2}}
\end{figure}

\section{The charged kaon form factor} 
Similar to the proton FF, the charged kaon TL FF was measured
at BABAR  with LA and SA  ISR techniques, allowing to study 
different $K^+K^-$ mass ranges. In the first LA measurement \cite{kk1}    
the  $K^+K^-$ mass range was studied  from  threshold up to 5 GeV/c$^2$.
This is the most precise measurement of the $e^+e^-\to K^+K^-$ process
below 2.6 GeV/c$^2$. 
The obtained FF  values (Fig.\ref{kkform1} (left)) are  about 4-5 times 
higher then  the asymptotic QCD prediction. 
\begin{figure}
\includegraphics[width=0.47\textwidth]{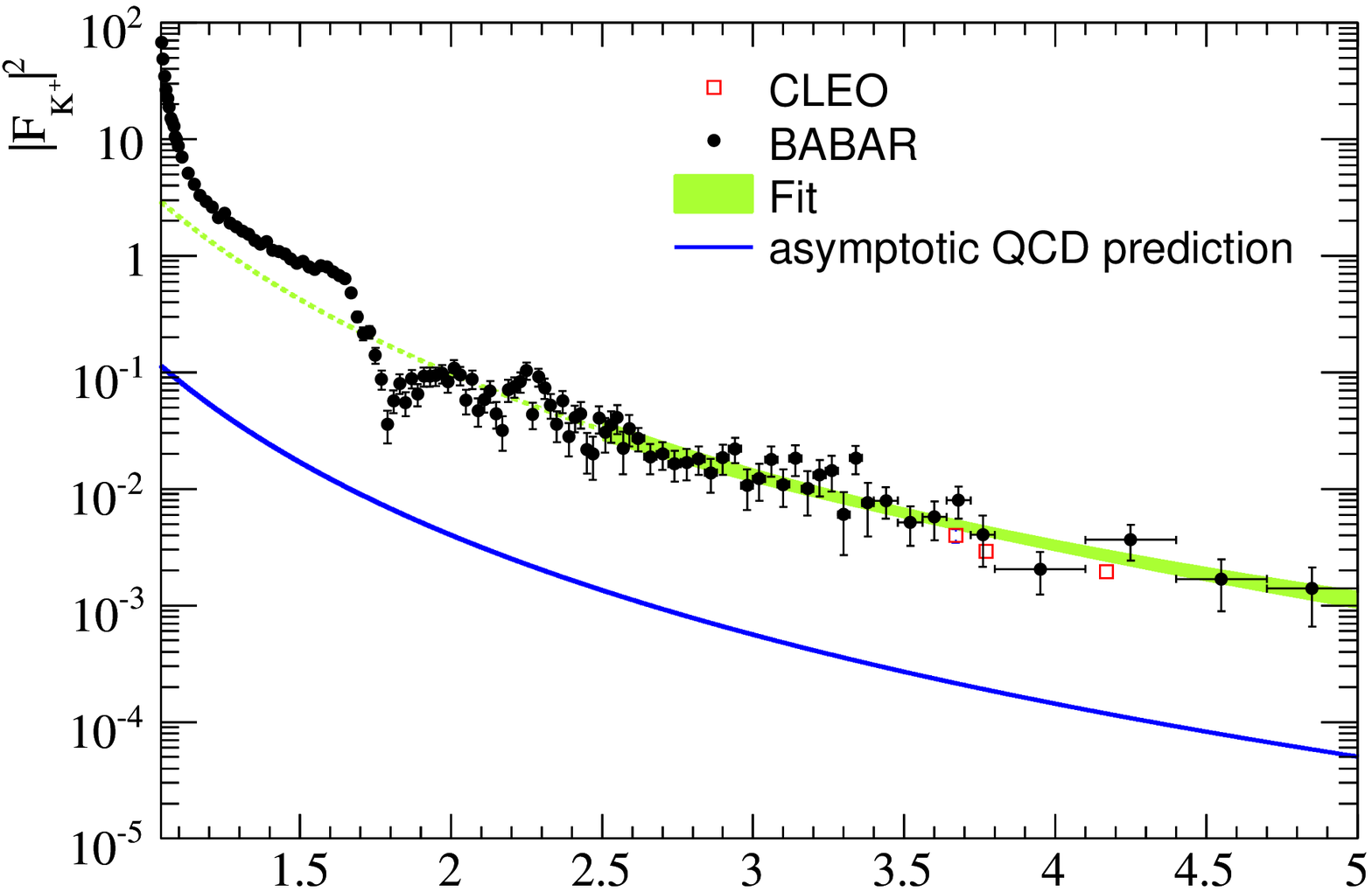}
\includegraphics[width=0.47\textwidth]{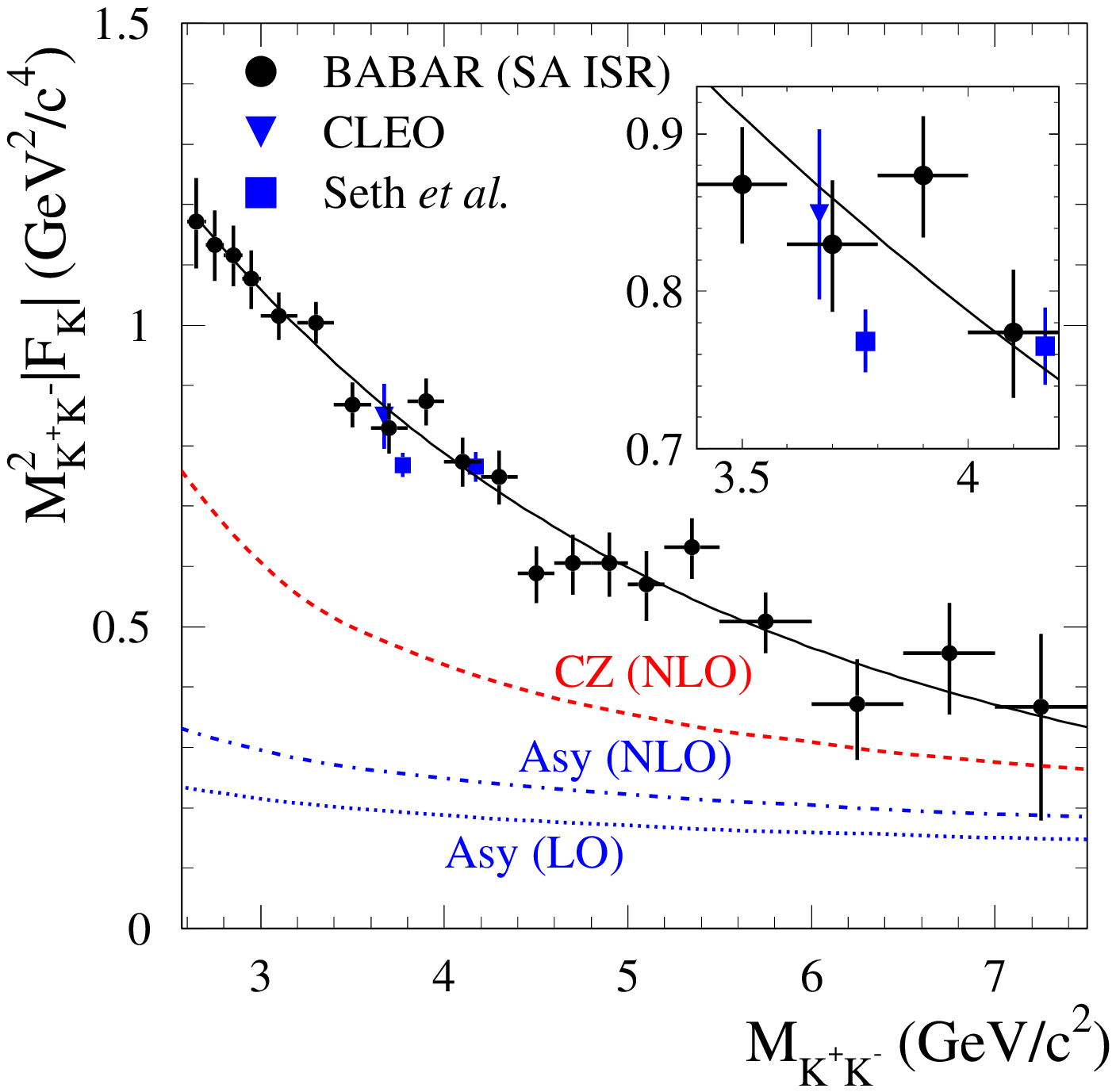}
\caption{ (Color online) {\bf Left}: 
The charged kaon form factor versus $M_{K^+K^-}$(GeV/c$^2$)
measured by BABAR  \cite{kk1}. {\bf Right}: 
The scaled charged kaon form factor measured by BABAR  \cite{kk2}.
Different QCD based  predictions are shown by the curves lying  
below data points. The region near $\Psi(3770)$ is shown in the inset.
\label{kkform1}}
\end{figure}

   In the second kaon FF measurement \cite{kk2} using the SA ISR 
technique,  the  $K^+K^-$ mass range was extended up  to 7.5  GeV/c$^2$.
The measured scaled kaon FF  ($M_{K^+K^-}^2F_K$) in the range 2.6-7.5 GeV/c$^2$
is shown in Fig.\ref{kkform1} (right). The QCD model prediction from  different
authors are shown by the curves lying below the experimental points. The 
references for these predictions can be found in the original BABAR paper 
\cite{kk2}. One can see in  Fig.\ref{kkform1} (right)   that
beginning from 6 GeV/c$^2$ the BABAR experimental point errors 
begin  to  intersect the QCD prediction. 
The main conclusion  from these data is that the kaon FF
begins  to approach to the  QCD asymptotic limit when energy increases. 

\section{Summary and Outlook}
In recent years, thanks to development of ISR technique
at  BABAR, a 
great progress has been achieved in the experimental study of TL 
electromagnetic FFs  of charged hadrons. In this talk new data
are presented for proton and  charged kaon FF. 
A comparison with the QCD  predictions is made. It was
found that at energy higher than 5 GeV for proton and 6  GeV
for the charged kaon the measured FF values begin to approach
to their asymptotic QCD limit.

{\bf Acknowledgment}. The author expresses his gratitude to
V.P. Druzhinin for fruitfull discussion.
 This work is partially
 supported in the framework of the State order of the Russian Ministry of
 Science and Education and the RFBR grants No. 15-02-01037 and Sci.School
 2479.2014.2.

\end{document}